\def\TMACLIB{}
\input \TMACLIB Macros.ttx
\input\TMACLIB stdbra.ttx 
\Refstda\Citebrackets 
\documentclass[10pt,twocolumn]{article}
\usepackage{newlfont}
 \usepackage{amsmath}
\usepackage{amssymb}
\usepackage{fixup}
\usepackage{stmaryrd}
\usepackage{program}
\usepackage{graphicx}

\MathsModeStringsfalse
\sfvariables

\newtheorem{theorem}{Theorem}[section] 
\newtheorem{defn}[theorem]{Definition}

\makeatletter
\newcommand{\FermaT}{Ferma\kern-0.125emT}

\newcommand{\MYELSE}{\@marginspace
                                                            \untab\quad\tab\kern-1em\hbox{\keyword{else}\ }}
\newcommand{\MYTHEN}{\@marginspace
                                                            \untab\quad\tab\kern-1em\hbox{\keyword{then}\ }}
\def\EQ{\EQspace\equiv\EQspace\\\quad\tab}
\let\@oldEQ=\EQ
\def\PROC{\quad\tab\kern-1em\keyword{proc}\ 
            \gdef\EQ{\ \equiv\ \global\let\EQ=\@oldEQ}}
\def\WHILE{\quad\tab\kern-1em\keyword{while}\ 
           \gdef\DO{\@marginspace\keyword{do}\ \global\let\DO=\@oldDO}}
\def\VAR{\quad\tab\kern-1em\keyword{var}\ }
\def\bigstrut{\rule[-1.2ex]{0pt}{3.8ex}}
\def\spaceabove{\rule{0pt}{2.6ex}}
\def\spacebelow{\rule[-1.2ex]{0pt}{0pt}}
\makeatother

\makeatletter
\def\@normalsize{\@setsize\normalsize{12pt}\xpt\@xpt
\abovedisplayskip 10pt plus2pt minus5pt\belowdisplayskip \abovedisplayskip
\abovedisplayshortskip \z@ plus3pt\belowdisplayshortskip 6pt plus3pt
minus3pt\let\@listi\@listI} 

\def\subsize{\@setsize\subsize{12pt}\xipt\@xipt}

\def\section{\@startsection {section}{1}{\z@}{24pt plus 2pt minus 2pt}
{12pt plus 2pt minus 2pt}{\large\bf}}

\def\subsection{\@startsection {subsection}{2}{\z@}{12pt plus 2pt minus 2pt}
{12pt plus 2pt minus 2pt}{\subsize\bf}}
\makeatother

\begin{document}

\date{}

\title{\Large\bf Reverse Engineering from Assembler to Formal Specifications
via Program Transformations}

\author%
{M.~P.~Ward \\
Software Technology Research Lab \\
De Montfort University\\
The Gateway,\\
Leicester LE1 9BH, UK\\
{\tt Martin.Ward@durham.ac.uk}}

\maketitle

\thispagestyle{empty}

\subsection*{\centering Abstract}
The \FermaT{} transformation system, based on research carried out over the last sixteen years at Durham
University, De Montfort University and Software Migrations Ltd., is an industrial-strength formal
transformation engine with many applications in program comprehension and language migration. This
paper is a case study which uses automated plus manually-directed transformations and abstractions to
convert an IBM 370 Assembler code program into a very high-level abstract specification.

\section{Introduction}

\noindent Keywords: Assembler, Migration, Comprehension, Formal Methods, Abstraction, WSL, Wide
Spectrum Language, Program Transformation, Legacy Systems, Restructuring.

\section{Introduction}

There is a vast collection of operational software systems which are vitally important to their users, yet
are becoming increasingly difficult to maintain, enhance and keep up to date with rapidly changing
requirements. For many of these so called \emph{legacy systems} the option of throwing the system
away and re-writing it from scratch is not economically viable. Methods are therefore urgently required
which enable these systems to evolve in a controlled manner. In particular, legacy assembler systems
have high maintenance costs, and migrating such systems to a different environment (eg.~a client-server
architecture) is much more difficult than for high-level language systems. The \FermaT{} transformation
system uses formal proven program transformations, which preserve or refine the semantics of a program
while changing its form. These transformations are applied to restructure and simplify the legacy systems
and to extract higher-level representations.

By using an appropriate sequence of transformations, the extracted representation is guaranteed to be
equivalent to the original code logic. The method is based on a formal wide spectrum language, called
WSL, with accompanying formal method. Over the last sixteen years we have developed a large
catalogue of proven transformations, together with mechanically verifiable applicability conditions.
These have been applied to many software development, reverse engineering and maintenance problems.

\section{Theoretical Foundation}

The theoretical work on which \FermaT{} is based originated not in software maintenance, but in research
on the development of a language in which proofs of equivalence for program transformations could be
achieved as easily as possible for a wide range of constructs.

WSL is the ``Wide Spectrum Language'' used in our program transformation work, which includes
low-level programming constructs and high-level abstract specifications within a single language. This
has the advantage that one does not need to differentiate between programming and specification
languages: the entire transformational development of a program from abstract specification to detailed
implementation can be carried out in a single language. Conversely, the entire reverse-engineering
process, from a transliteration of the source program to a high-level specification, can also be carried
out in the same language. During either of these processes, different parts of the program may be
expressed at different levels of abstraction. So a wide-spectrum language forms an ideal tool for
developing methods for formal program development and also for formal reverse engineering (for which
we have coined the term \emph{inverse engineering}).

A \emph{program transformation} is an operation which modifies a program into a different form
which has the same external behaviour (i.e.~it is equivalent under a precisely defined denotational
semantics). Since both programs and specifications are part of the same language, transformations can
be used to demonstrate that a given program is a correct implementation of a given specification.

A \emph{refinement} is an operation which modifies a program to make its behaviour more defined
and/or more deterministic. A typical implementation of a nondeterministic specification will be a
\emph{refinement} rather than a strict equivalence. The opposite of refinement is \emph{abstraction}:
we say that a specification is an abstraction of a program which implements it. See\Lspace \Lcitemark 5\Citecomma
6\Rcitemark \Rspace{} and\Lspace \Lcitemark 1\Rcitemark \Rspace{} for a description of refinement.

The syntax and semantics of WSL are described in\Lspace \Lcitemark 8\Citecomma
9\Citecomma
12\Rcitemark \Rspace{} so will not be discussed in detail here. Most of the constructs in WSL, for example
\keyword{if} statements, \keyword{while} loops, procedures and functions, are common to many
programming languages. However there are some features relating to the ``specification level'' of the
language which are unusual. Expressions and conditions (formulae) in WSL are taken directly from
first order logic: in fact, an infinitary first order logic is used (see\Lspace \Lcitemark 4\Rcitemark \Rspace{} for details), which
allows countably infinite disjunctions and conjunctions. This use of first order logic means that
statements in WSL can include existential and universal quantification over infinite sets, and similar
(non-executable) operations. Two list operators are also used in specifications: for a unary function
$f$ and list $L=\langle a_{1}, \dots , a_{n}\rangle $ the map operator $\bstar $ is defined:
\[
 f\bstar L \edf  \langle f(a_{1}), f(a_{2}), \dots , f(a_{n})\rangle 
\]
For a binary operator $g$ and non-empty list $L$ the reduce operator $\bdiv $ is defined:
\begin{eqnarray*}
g\bdiv L        & \edf  &       a_{1} \qquad\text{if $n = 1$} \\
                & \edf  &       g(a_{1}, g\bdiv \langle a_{2}, \dots , a_{n}\rangle )   \quad\text{if $\ell (L) > 1$}
\end{eqnarray*}
For example, if $f$ is a function which returns integers, and $L$ is a non-empty list of suitable
arguments for $f$, then ${+}\bdiv f\bstar L$ is the result of applying $f$ to every element of $L$ and adding up
the results. We also use $\ell (L)$ to denote the length of list $L$ and $L[i\rdots j]$ to denote the sublist
$\langle a_{i}, \dots , a_{j}\rangle $.

Over the last sixteen years we have been developing the WSL language, in parallel with the development
of a transformation theory and proof methods. Over this time the language has developed from a simple
and tractable kernel language\Lspace \Lcitemark 9\Rcitemark \Rspace{} to a complete and powerful programming language. At the
``low-level'' end of the language there exist automatic translators from IBM Assembler into WSL, and
from a subset of WSL into C. At the ``high-level'' end it is possible to write abstract specifications,
similar to {\bf Z} and VDM.

In\Lspace \Lcitemark 10\Citecomma
13\Rcitemark \Rspace{} program transformations are used to derive a variety of efficient
algorithms from abstract specifications. In\Lspace \Lcitemark 10\Citecomma
12\Citecomma
13\Rcitemark \Rspace{} the same
transformations are used in the reverse direction: using transformations to derive a concise abstract
representation of the specification for several challenging programs.

In\Lspace \Lcitemark 11\Rcitemark \Rspace{} we describe a case study using \FermaT{} to migrate an assembler program to modular
and maintainable C code, using purely automatic transformations with no manual intervention. As far as
we know, none of the other researchers in program transformations (for example,\Lspace \Lcitemark 2\Citecomma
7\Rcitemark \Rspace{}) have attempted to apply their methods to assembler code. The nearest research is Cristina
Cifuentes work on decompilation and binary translation\Lspace \Lcitemark Cifuentes CSMR00\Rcitemark \Rspace{}.

In this paper we go even further in the reverse engineering process. Starting with the same assembler
program from\Lspace \Lcitemark 11\Rcitemark \Rspace{} we use formal transformations to abstract an equivalent high-level
specification of the program.

\section{Example Transformations in \FermaT{}} \label{sec-examples}

In this section we describe a small number of the transformations implemented in \FermaT{} which are
used in the case study. If \S{1} and \S{2} are any WSL statements and $\Delta $ is any countable set of formulae
with no free variables, then we write $\Delta \vdash \S{1}\LE \S{2}$ to denote that \S{2} is a refinement of \S{1} whenever all
the formulae on $\Delta $ are true. If $\Delta \vdash \S{1}\LE \S{2}$ and $\Delta \vdash \S{2}\LE \S{1}$ then we write $\Delta \vdash \S{1}\EQT \S{2}$ and say
that \S{1} and \S{2} are \emph{equivalent}. If \S{2} is generated from \S{1} by a program transformation, then
$\Delta \vdash \S{1}\EQT \S{2}$, where $\Delta $ is the set of applicability conditions for the transformation.

\subsection{Expand Forwards}

If \B{} is any condition and \S{1}, \S{2} and \S{3} are any statements then:
\begin{multline*}
                               \Delta \vdash  \IF  \B{} \THEN  \S{1} \ELSE  \S{2} \FI ; \S{3} \EQT \\  \IF  \B{} \THEN  \S{1}; \S{3} \ELSE  \S{2};\S{3} \FI 
\end{multline*}

\subsection{Loops}

As well as the usual $\FOR $ and $\WHILE $ loops, there is a notation for unbounded loops. Statements of
the form $\DO  \S{} \OD  $, where \S{} is a statement, are ``infinite'' or ``unbounded'' loops which can only be
terminated by the execution of a statement of the form $\EXIT (n)$ which causes the program to exit the
$n$ enclosing loops. We use $\EXIT $ as an abbreviation for $\EXIT (1)$. To simplify the language we
disallow \keyword{exit}s which leave a block or a loop other than an unbounded loop. We also insist
that $n$ be an integer, \emph{not} a variable or expression---this ensures that we can always
determine the target of the $\EXIT $.

\begin{defn} Global Substitution\\\normalshape
If $\P{}(\S{}, p)$ is a predicate on a statement \S{} and position $p$ within \S{}, and $\S'(\S{}, p)$ is a function
which returns a statement for any given statement \S{} and $p$, then the effect of replacing or appending to
the statement at position $p$ in \S{} with $\S'(\S{}, p)$ for every $p$ such that $\P{}(\S{}, p)$ holds is denoted:
\[
                                                              \S{}\sub[\S'(\S{}, p)/p|\P{}(\S{}, p)]
\]
\end{defn}
If the statement at position $p$ in \S{} is an $\EXIT $ statement, then it is replaced by $\S'(\S{}, p)$.
Otherwise, $\S'(\S{}, p)$ is appended in sequence after the statement at position $p$.

Within a global substitution we use $\delta (\S{}, p)$ to denote the \emph{depth} of a component of a
statement. This is the number of enclosing $\DO \dots \OD $ loops surrounding the component. We use $\tau (\S{},
p)$ to denote the terminal value of a statement. This is the number of enclosing loops around \S{} which
might be terminated by execution of the statement at position $p$ in \S{}. If the statement at position $p$
in \S{} does not terminate \S{} then $\tau (\S{}, p) = -1$. For example, any $\EXIT (n)$ has terminal value $n$. If
\S{} contains an $\EXIT (n)$ within $m$ nested loops (where $m \leqslant  n$) then the terminal value of \S{} itself,
denoted $\tau (\S{}, \langle \rangle )$, will be at least $n-m$. A statement \S{} with terminal value zero cannot terminate
any enclosing loops, so the next thing to be executed after \S{} will be the next statement in the sequence
containing \S{} (if there is one). Such a statement is called a \emph{proper sequence}. If \S{} is a proper
sequence, then:
\[
                                         \Delta \vdash  \DO  \IF  \B{} \THEN  \EXIT  \FI ; \S{} \OD  \EQT  \WHILE  \NOT \B{} \DO  \S{} \OD 
\]

In the following transformations, the global substitutions are all applied to the \emph{simple terminal
statements} of \S{}. These are the statements which are neither a sequence, a conditional, or a $\DO \dots \OD $
loop and which will terminate \S{} if they are executed. For example, in:
\begin{program}
\IF  \B{} \THEN  x:= 1; y := 2 \ELSE  \EXIT  \FI 
\end{program}
the terminal statements are $y := 2$ and $\EXIT $. If the statement is enclosed in a $\DO \dots \OD $ loop, only
the $\EXIT $ will be a terminal statement.

We usually omit the parameters from $\delta $ and $\tau $ in a global substitution when these are obvious from
the context.

\begin{defn} Incrementation\\\normalshape The incrementation of \S{} by $n$ (where $n$ is any
non-negative integer) is defined as the incrementation of all simple terminal statements in \S{}. An
$\EXIT $ is incremented by incrementing its parameter, while any other simple statement is incremented by
appending an $\EXIT $:
\[
                                                     \S{}+n \edf  \S{}\sub[\EXIT (n+\delta )/p|\tau \geqslant 0]
\]
For example:
\begin{multline*}
\IF  \B{} \THEN  x:= 1; y := 2 \ELSE  \EXIT  \FI  + 2 \\
{} = \IF  \B{} \THEN  x:= 1; y := 2; \EXIT (2) \ELSE  \EXIT (3) \FI 
\end{multline*}
while:
\begin{multline*}
\DO  \IF  \B{} \THEN  x:= 1; y := 2 \ELSE  \EXIT  \FI  \OD  + 2 \\
{} = \DO  \IF  \B{} \THEN  x:= 1; y := 2 \ELSE  \EXIT (3) \FI  \OD 
\end{multline*}
\end{defn}

\begin{defn} Partial Incrementation\\\normalshape
The notation $\S{}+(n, m)$ where $m\geqslant 0$ denotes incrementation of the terminal statements in \S{} with
terminal value $m$ or greater:
\[
                                                \S{}+(n, m) \edf  \S{}\sub[\EXIT (n+\delta )/p|\tau  \geqslant m]
\]
\end{defn}
Note that $\DO  \S{} \OD +(n, m) = \DO  \S{}+(n, m+1) \OD $.

\subsection{Absorption}

For any statements \S{1} and \S{2}:
\[
                                                      \Delta \vdash  \S{1}; \S{2} \EQT  \S{1}\sub[\S{2}+\delta /p|\tau =0]
\]
For example:
\begin{multline*}
\DO  \IF  \B{} \THEN  x:= 1; y := 2 \ELSE  \EXIT  \FI  \OD ; z := 1 \\
\DO  \IF  \B{} \THEN  x:= 1; y := 2 \ELSE  z := 1; \EXIT  \FI  \OD  \\
\end{multline*}
This transformation can be applied in reverse to ``take out'' code from a loop.

\subsection{False Loop}

We can insert a loop around any statement, by incrementing it first:
\[
                                                                     \Delta \vdash \S{} \EQT  \DO  \S{}+1 \OD 
\]
(This is a ``false loop'' because the body of the loop can only be executed once).

\subsection{Loop Doubling}\label{sec-loop-doubling}

Any loop can be converted to a double loop by the last transformation, or by incrementing the body of
the loop:
\begin{eqnarray*}
\Delta \vdash  \DO  \S{} \OD    & \EQT  &        \DO  \DO  \S{} \OD +1 \OD  \\
                                & \EQT  &        \DO  \DO  \S{}+1 \OD  \OD 
\end{eqnarray*}
More generally, we can arbitrarily decide whether or not to increment each terminal statement in \S{} with
terminal value zero:
\begin{program}
\Delta \vdash  \DO  \S{} \OD  \EQT  {}
\DO  \DO  \S{}\,[\,\tab \EXIT (\delta +1)/p
                     \,\origbar\, \tau >0 \OR  \tau =0 \AND  \Psi (\S{}, p)\,]\, \OD  \OD 
\end{program}
Where $\Psi $ is \emph{any} condition on \S{} and $p$.

This can be combined with the inverse of absorption to ``isolate'' part of a loop body. For example:
\begin{program}
\Delta \vdash \DO  \S{}; \IF  \B{} \THEN  \S{1} \ELSE  \S{2} \FI  \OD 
{}\EQT  \tab     \DO  \DO       \S{}+(1, 1);
                                        \IF  \B{} \THEN  \EXIT  \ELSE  \S{2}+(1, 1) \FI  \OD ;
                                \S{1} \OD 
\end{program}

\subsection{Loop Inversion}

If \S{1} is a proper sequence then:
\[
                                                      \Delta \vdash  \DO  \S{1}; \S{2} \OD  \EQT  \S{1}; \DO  \S{2}; \S{1} \OD 
\]
More generally, for \emph{any} statements \S{1} and \S{2}:
\[
                                        \Delta \vdash  \DO  \S{1}; \S{2} \OD  \EQT  \DO  \S{1}; \DO  \S{2}; \S{1} \OD +1 \OD 
\]

\subsection{Loop Unrolling}

We can unroll the first step of a loop:
\begin{eqnarray*}
\Delta \vdash  \DO  \S{} \OD  \EQT  \S{} &&\sub[\DO  \S{} \OD +\delta +1/p|\tau =0]\\
                                        &&\sub[\EXIT (\tau +\delta -1)/p|\tau \geqslant 1]
\end{eqnarray*}
where the RHS contains two successive global substitutions on \S{}.

More generally, we can insert a copy of the whole loop, with certain terminal statements of the loop body
incremented, after certain terminal statements in the loop body. Let \S' be formed from \S{} by
incrementing selected terminal statements with terminal value zero:
\[
                                                 \S' = \S{}\sub[\EXIT (\delta +1)/p|\tau =0 \AND  \Phi (\S{}, p)]
\]
where $\Phi $ is any condition (see Section~\ref{sec-loop-doubling}). Then:
\begin{eqnarray*}
&& \Delta \vdash  \DO  \S{} \OD  \\
&& {}\EQT       \DO  \S{} \begin{aligned}[t] &\sub[\DO  \S' \OD +\delta +1/p|\tau =0 \AND  \Psi (\S{}, p)]\\
                                                                                &\sub[\EXIT (\tau +\delta -1)/p|\tau \geqslant 1] \OD  \end{aligned}
\end{eqnarray*}
where $\Psi $ is any condition.

\section{Modelling Assembler in WSL} \label{sec-modelling}

Constructing a useful scientific model necessarily involves throwing away some information: in other
words, to be useful a model \emph{must} be inaccurate, or at least idealised, to a certain extent. For
example ``ideal gases'', ``incompressible fluids'' and ``billiard ball molecules'' are all useful models
which gain their utility by abstracting away some details of the real world. In the case of modelling a
programming language, such as Assembler, it is theoretically possible to have a perfect model of the
language which correctly captures the behaviour of all assembler programs. Certain features of
Assembler, such as branching to register addresses, self-modifying code and so on, would imply that
such a model would have to record the entire state of the machine, including all registers, memory, disk
space, and external devices, and ``interpret'' this state as each instruction is executed. (Consider the
effect of loading some data from a disk file into memory, performing arithmetic at arbitrary places in the
data, and then branching to the start of the data block!) Unfortunately, such a model is useless for
reverse engineering or migration purposes.

What we need is a practical model for assembler programs which is suitable for reverse engineering, and
is accurate enough to deal with all the programming constructs which are likely to be encountered.

\subsection{Assembler to WSL Translation}

The aim of the assembler to WSL translator is to generate WSL code which models as accurately as
possible the behaviour of the original assembler module, without worrying too much about the size,
efficiency or complexity of the resulting code. Typically, the raw WSL translation of an assembler
module will be three to five times bigger than the source file and have a very high McCabe cyclomatic
complexity (typically in the hundreds, often in the thousands). This is, in part, because every ``branch to
register'' instruction branches to the dispatch routine, which in turn contains branches to every possible
return point. In addition, every instruction which sets the ``condition code'' flags will is translated into
WSL code which assigns an appropriate value to a special variable |cc| (to emulate the condition code):
whether or not the condition code is subsequently tested. See\Lspace \Lcitemark 11\Rcitemark \Rspace{} for further details of the
assembler to WSL translation process and the various features of commercial assembler code which it
has to deal with.

However, the \FermaT{} transformation engine includes some very powerful transformations for
simplifying WSL code, removing redundancies, tracking dispatch codes, and so on. In most cases \FermaT{}
can automatically unscramble the tangle of ``branch and save'' and ``branch to register'' code to extract
self-contained, single-entry single-exit procedures and so eliminate the dispatch procedure. In
addition, \FermaT{} can nearly always eliminate the |cc| variable by constructing appropriate conditional
statements.

\section{The Sample Program}

Our sample program is from ``A Guided Tour of Program Design Methodologies'', by G.~D.~Bergland
\Lcitemark 3\Rcitemark \Rspace{} who in turn took it from a story called ``Getting it Wrong'' that has been related by
Michael Jackson on numerous occasions:
\begin{program}
\PROC  |Management_Report| \EQ 
        \VAR  |SW1|:=0, |SW2|:=0:
                |Produce_Heading|;
                |read|(|stuff|);
                \WHILE  |NOT|~|eof|(|stuff|) \DO 
                        \IF  |First_Record_In_Group|
                          \THEN  \IF  |SW1|=1
                                          \THEN  |Process_End_Of_Previous_Group|
                                        \FI ;
                                        |SW1|:=1;
                                        |Process_Start_Of_New_Group|;
                                        |Process_Record|;
                                        |SW2|:=1
                          \MYELSE 
               |Process_Record|; |SW2|:=1
                        \FI ;
                        |read|(|stuff|)
                \OD ;
                \IF  |SW2|=1 \THEN  |Process_End_Of_Last_Group|
                \FI ;
                |Produce_Summary|
        \END \fullstop 
\end{program}
The program is a simple report generator which reads a sorted transaction file: each transaction contains
the name of an item and the amount received or distributed from the warehouse. The program generates
a report showing the net change in inventory for each item in the transaction file.

Our resident assembler guru was given the above pseudocode and asked to write an assembler
implementation which uses as many ``features'' of assembler as possible. The result is given in
Section~\ref{sec-src} (I should like to point out on his behalf that this is \emph{not} his normal
coding style!) The program includes self-modifying code: the ``first time through switch'' \texttt{SW1} is
implemented by modifying the branch labelled \texttt{LAAA} to a \texttt{NOP} in the instruction labelled \texttt{LAB}, and an
EXecute statement has been used to get a variable length move.

\section{Automatic Program Transformation}

The first stage in the transformation process is Data Translation. This transformation uses the
restructured data file to change the data representation in the program. Initially all data is accessed
directly from memory (represented as the byte array |a|) by adding the base register to the displacement
to get an address. The restructured data file gives the layout of all data in memory, so by making some
reasonable assumptions about non-overlapping DSECTS etc., \FermaT{} is able to transform the program
into an equivalent program where the data is accessed directly through variables and structures. For
example, consider the ``raw WSL'' statement:
\begin{program}
|!P|~|mvc|(\tab |a|[|db|(|writem|, |r3|), 3 + 1]
                          \var{} |a|[|db|(|wlast|, |r3|), 3 + 1]);\untab
\end{program}
Here, the |!P| indicates an external procedure call to the |mvc| procedure which implements the MVC
(move characters) instruction. This moves the given number of characters from the given source address
to the given destination address. The function $|db|(x, y)$ simply returns $x+y$, the displacement plus
the base register, so the source address is $|writem|+|r3|$ and the destination address is $|wlast|+|r3|$.
After data translation, the same names are used as the actual variables and the base registers are
eliminated.

This statement is automatically transformed into the simple assignment:
\begin{program}
|wlast| := |wrec|.|writem|;
\end{program}
In the case of our simple program, there is only one structure to uncover: the |wrec| print record
which contains fields |writem|, |wrtype| and |wrqty| plus some unnamed fillers.

The next stage is control flow restructuring: eliminating non-essential labels and branches, introducing
loops. This is carried out in a series of passes through the program, at each iteration the program is
searched for points where a simplifying transformation (such as loop insertion or branch merging) can
be applied. The iteration is continued until no further improvement can be achieved.

The raw WSL is written as an \emph{action system}, a collection of parameterless procedures (actions)
where execution of any actuin will always lead to either calling another action, or calling the special
action $Z$ which terminates the whole action system. An action system itself is a simple statement, so
action systems can be nested inside each other, but a sub-action system cannot call actions in the main
system.

The system then analyses the remaining actions to determine which actions may form the body of a
simple procedure. To do this it uses both control flow and data flow analysis. If it determines that a
collection of actions form a procedure, then these actions are extracted out as a sub-action system in the
body of the procedure.

After control flow restructuring we have data flow analysis: in particular an extended form of constant
propagation which can propagate return addresses through procedure calls. If a |dispatch| call is
encountered with a known |destination| value, then it can be unfolded and simplified. The same
transformation also deals with conditional assignments to the condition code (|cc|) in order to remove
references to |cc| where possible.

\FermaT{} was able to extract a collection of actions to form the |endgroup| procedure, so that the code:
\begin{program}
|r10| := 112; \CALL  |endgroup|
\end{program}
becomes:
\begin{program}
|r10| := 112; |endgroup|(); \CALL  |dispatch|
\end{program}
\FermaT{} determines that the value in |r10| will be copied into |destination| by the body of |endgroup|.
Within |dispatch| the value in |destination| is compared against the offsets of all the possible return
points. Offset 112 is associated with the label |lab|, so this $\CALL  |dispatch|$ can be replaced by $\CALL 
|lab|$.

The control flow and data flow restructuring transformations are iterated until no further improvement
is possible. Figure~\ref{fig-metrics} lists the metrics for the raw WSL translation and after automatic
restructuring and simplifying transformations have been applied. This order of magnitude improvement
in most of the metrics is typical for all sizes of assembler module. See\Lspace \Lcitemark 11\Rcitemark \Rspace{} for more
details of this part of the transformation process.

\def\mline{\unskip\spacebelow\\\hline\spaceabove}
\begin{figure}[htbp]
\begin{center}
\begin{tabular}{|lrr|}
\hline\bigstrut
Metric                  & \llap{Raw WSL}        & Structured WSL
\mline
Statements                              & 561                   & 106                   \\
Expressions                     & 1,589         & 210                   \\
McCabe                          & 184                   & 17                    \\
Control/Data Flow       & 520                   & 156                   \\
Branch--Loop            & 145                   & 17                    \\
Structural                              & 6,685         & 75%
1\spacebelow\\
\hline
\end{tabular}
\caption{Metrics Before and After Transformation}\label{fig-metrics}
\end{center}
\end{figure}

\begin{program}
\BEGIN  
  |f_laaa| := 1;
  |!P|~|open|(|ddin_ddname|, |input| \var{} |os|);
  |!P|~|open|(|rdsout_ddname|, |output| \var{} |os|);
  |wprt|[1\rdots 17] := ``MANAGEMENT REPORT'';
  |write1|(); |write1|();
  |wprt|[1\rdots 20] := ``ITEM      NET CHANGE'';
  |write1|(); |write1|();
  |xsw1| := 0;
  \DO  |r0| := 0; |r1| := 0; |r15| := 0;
     |!P|~|get|(|ddin_ddname| \var{} |os|, |r0|, |r1|, |r15|, |wrec|);
     \IF  |!XC|~|end_of_file|(|ddin_ddname|)
       \THEN  \EXIT (1) \FI ;
     \IF  |wrec|.|writem| \neq |wlast|
       \THEN  \IF  |f_laaa| \neq 1
              \THEN  |endgroup|() \FI ;
            |f_laaa| := 0;
            |wlast| := |wrec|.|writem|;
            |wnet| := |!XF|~|zap|(``hex 0x0C'') \FI ;
     |worka| := |!XF|~|pack|(|wrec|.|wrqty|, 2);
     \IF  |wrec|.|wrtype| \neq ``R''
       \THEN  |wnet| := |!XF|~|sp|(|wnet|, |worka|)
       \ELSE  |wnet| := |!XF|~|ap|(|wnet|, |worka|) \FI ;
     |xsw1| := ``hex 0xFF'' \OD ;
  \IF  |xsw1| = ``hex 0xFF'' \THEN  |endgroup|() \FI ;
  |wprt|[1\rdots 17] := ``NUMBER CHANGED = '';
  |!P|~|ed|(|wchange|[1\rdots 10] \var{} |workb|);
  |r4| := |!XF|~|address_of|(|workb|); |r1| := 9;
  \DO  \IF  a[|r4|, 1] \neq `` ''  \THEN  \EXIT (1) \FI ;
     |r4| := |r4| + 1;
     |r1| := |r1| - 1;
     \IF  |r1| = 0 \THEN  \EXIT (1) \FI  \OD ;
  a[|!XF|~|address_of|(|wprt|) + 17, |r1| + 1]
  \quad := a[|r4|, |r1| + 1];
  |write1|();
  |!P|~|close|(|ddin_ddname| \var{} |os|);
  |!P|~|close|(|rdsout_ddname| \var{} |os|)
\WHERE  
   \PROC  |endgroup|() \EQ 
      |wprt|[1\rdots 4] := |wlast|;
      |wsign| := ``+'';
      \IF  |wnet| < ``hex 0x0C''  \THEN  |wsign| := ``-'' \FI ;
      |wprt|[8\rdots 17] := ``hex 0x40206B2020206B202120'';
      |!P|~|edmk|(|wnet|[1\rdots 10] \var{} |wprt|[8\rdots 17], |r1|);
      |r1| := |r1| - 1;
      a[|r1|, 1] := |wsign|;
      |write1|(); |write1|();
      |wchange| := |!XF|~|ap|(|wchange|, ``hex 0x1C'') \END ,
  \PROC  |write1|() \EQ 
      |!P|~|put|(|rdsout_ddname|, |wprt| \var{} |os|);
      |wprt| := |wspaces| \END 
\END 
\end{program}

\section{Abstracting a Specification}

This is about as far as the \FermaT{} system can get by purely automatic transformation applications with
no human intervention. The next step in the abstraction process is to change the data representation so
that files become lists. We unfold the |write1| procedure and replace |zap|, |ap| and |sp| calls by
their actual operations. We abstract away from the layout of the output file by creating a list of the data
elements that appear on each line of output and appending this list to the |output| array:
\begin{program}
\BEGIN 
  i := 0; |f_laaa| := 1;
  |output| := \langle \tab\langle ``MANAGEMENT REPORT''\rangle ,
                             \langle ``ITEM~~~~~~NET CHANGE''\rangle \rangle ;\untab
  |xsw1| := 0;
  \DO  i := i + 1; |wrec| := |input|[i];
        \IF  i \geqslant n \THEN  \EXIT (1) \FI ;
        \IF  |wrec|.|writem| \neq |wlast|
          \THEN  \IF  |f_laaa| \neq 1 
                          \THEN  |endgroup|() \FI ;
                    |f_laaa| := 0;
                    |wlast| := |wrec|.|writem|;
                    |wnet| := 0 \FI ;
        \IF  |wrec|.|wrtype| \neq ``R''
          \THEN  |wnet| := |wnet| - |wrec|.|wrqty|
          \ELSE  |wnet| := |wnet| + |wrec|.|wrqty| \FI ;
        |xsw1| := ``hex 0xFF'' \OD ;
  \IF  |xsw1| = ``hex 0xFF'' \THEN  |endgroup|() \FI ;
  |output| := \tab|output| \concat  {}
                                \quad\langle \langle ``NUMBER CHANGED = '', |wchange|\rangle \rangle ;\untab
\WHERE  
   \PROC  |endgroup|() \EQ 
       |output| := |output| \concat  \langle \langle |wlast|, |wnet|\rangle \rangle ;
      |wchange| := |wchange| + 1 \END 
\END 
\end{program}

We can get rid of the switches |xsw1| and |f_laaa| by unrolling the first
step of the $\DO \dots \OD $ loop and simplifying. We then use \emph{loop inversion}
to move some statements to the top of the loop:
\begin{program}
  i := i + 1; |wrec| := |input|[i];
  \IF  i \geqslant n
    \THEN  \SKIP{}
    \ELSE  |wlast| := |wrec|.|writem|;
         |wnet| := 0;
         \DO  \IF  |wrec|.|wrtype| \neq ``R''
              \THEN  |wnet| := |wnet| - |wrec|.|wrqty|
              \ELSE  |wnet| := |wnet| + |wrec|.|wrqty| \FI ;
            i := i + 1; |wrec| := |input|[i];
            \IF  |wrec|.|writem| \neq |wlast| \OR  i \geqslant n
              \THEN  |endgroup|();
                   \IF  i \geqslant n
                     \THEN  \EXIT (1)
                     \ELSE  |wlast| := |wrec|.|writem|;
                          |wnet| := 0 \FI  \FI  \OD  \FI ;
\end{program}

We want to roll the two statements $|LAST| := |wrec|.|writem|; |wnet| := 0$ into the top of the
loop, so convert the loop to a double-nested loop (\emph{loop doubling}) and take the statements out
of the inner loop (\emph{take out of loop}). Then apply \emph{loop inversion}. We can then take the
statements starting with $|endgroup|()$ out of the inner loop also:
\begin{program}
  i := i + 1; |wrec| := |input|[i];
  \IF  i \geqslant n
    \THEN  \SKIP{}
    \ELSE  \DO  |wlast| := |wrec|.|writem|;
            |wnet| := 0;
            \DO  \IF  |wrec|.|wrtype| \neq ``R''
                 \THEN  |wnet| := |wnet| - |wrec|.|wrqty|
                 \ELSE  |wnet| := |wnet| + |wrec|.|wrqty| \FI ;
               i := i + 1; |wrec| := |input|[i];
               \IF  |wrec|.|writem| \neq |wlast| \OR  i \geqslant n
                 \THEN  \EXIT (1) \FI  \OD ;
            |endgroup|();
            \IF  i \geqslant n \EXIT (1) \FI  \OD  \FI ;
\end{program}
Finally, the outer \keyword{if} statement can be removed by converting the outer loop
to a \keyword{while} loop (this is the \emph{floop to while} transformation):
\begin{program}
  i := i + 1; |wrec| := |input|[i];
  \WHILE  i < n \DO 
    |wlast| := |wrec|.|writem|;
    |wnet| := 0;
    \DO  \IF  |wrec|.|wrtype| \neq ``R''
                \THEN  |wnet| := |wnet| - |wrec|.|wrqty|
                \ELSE  |wnet| := |wnet| + |wrec|.|wrqty| \FI ;
         i := i + 1; |wrec| := |input|[i];
       \IF  |wrec|.|writem| \neq |wlast| \OR  i \geqslant n
         \THEN  \EXIT (1) \FI  \OD ;
    |endgroup|() \OD ;
\end{program}

Note that, after the initialisation code, the invariant $|wrec| = |input|[i]$ is always true, and for
$i>1$, $|wlast| = |input|[i-1].|writem|$ is also true, as is the invariant $|wchange| =
\ell (|output|)-2$. So we can remove these three variables from the program.

The program now consists of two simple nested loops, the outer \keyword{while} loop iterates
over the groups of records and ends with a call to $|endgroup|()$, while the inner $\DO \dots \OD $ loop
iterates over the records in the group.

This suggests that we restructure the data to more closely match the control structure of the program by
converting the input array to a list of lists where each sublist consists of a single group of data elements,
so that the outer loop processes sublists one at a time and the inner loop processes elements of each
sublist. The key to the data restructuring is to split the input sequence into sections such that the outer
loop processes one segment per iteration. This is easily achieved with a function $|split|(p, B)$
which splits $p$ into non-empty sections with the section breaks occurring between those pairs of
elements of $p$ where $B$ is false. (See\Lspace \Lcitemark 12\Rcitemark \Rspace{} for a formal definition of |split|).
In our case, the terminating condition on the inner loop provides the predicate on which to split:
\begin{program}
\FUNCT  |same_item|(x, y) \BODY
        x.|writem| = y.|writem|\fullstop 
\end{program}
Then the new variable $q$ is introduced with the assignment: $ q:=|split|(|input|, |same_item|)$.
We index the $q$ list with two variables $k_{1}$ and $k_{2}$ so that $q[k_{1}][k_{2}] = |input|[i]$. To do this
we preserve the invariant:
\[
i = {+}\bdiv (\ell \bstar q[1\rdots k_{1}-1])+k
\]
which, together with the invariant $|input| = {\concat }\bdiv q$ gives the required relationship. Adding these
ghost variables to the program we get: \begin{program}
q := |split|(|input|, |same_item|);
i := 1; k_{1}:= 1; k_{2} := 1;
  \WHILE  i < \ell (|input|) \DO 
    |wnet| := 0;
    \DO  \IF  |input|[i].|wrtype| \neq ``R''
                \THEN  |wnet| := |wnet| - |input|[i].|wrqty|
                \ELSE  |wnet| := |wnet| + |input|[i].|wrqty| \FI ;
       i := i + 1;
       k_{2} := k_{2} + 1;
       \IF  k_{2} > \ell (q[k_{1}]) \THEN  k_{1} := k_{1} + 1; k_{2} := 1 \FI ;
       \IF  \tab|input|[i].|writem| \neq |input|[i-1].|writem|
           {} \OR  i \geqslant \ell (|input|) \THEN  \EXIT (1) \FI  \OD ;
    |endgroup|() \OD ;
\end{program}
We can now replace references to the concrete variables |input| and $i$ by references to the new
variables $q$, $k_{1}$ and $k_{2}$. The key point is that $i<\ell (|input|)$ if and only if $k_{1}<\ell (q)$ and
\[
|input|[i].|writem| \neq |input|[i-1].|writem|
\]
is true when we have just moved into a new section of the input: in other words, precisely when $k_{2}=1$.
So we can remove the concrete variables from the program:
\begin{program}
q := |split|(|input|, |same_item|);
k_{1}:= 1; k_{2} := 1;
  \WHILE  k_{1} < \ell (q) \DO 
    |wnet| := 0;
    \DO  \IF  q[k_{1}][k_{2}].|wrtype| \neq ``R''
                \THEN  |wnet| := |wnet| - q[k_{1}][k_{2}].|wrqty|
                \ELSE  |wnet| := |wnet| + q[k_{1}][k_{2}].|wrqty| \FI ;
       k_{2} := k_{2} + 1;
       \IF  k_{2} > \ell (q[k_{1}]) \THEN  k_{1} := k_{1} + 1; k_{2} := 1 \FI ;
       \IF  k_{2}=1 \THEN  \EXIT (1) \FI  \OD ;
    |endgroup|() \OD ;
\end{program}
Now the inner loop reduces to a simple \keyword{for} loop:
\begin{program}
q := |split|(|input|, |same_item|);
k_{1}:= 1;
  \WHILE  k_{1} < \ell (q) \DO 
    |wnet| := 0;
    \FOR  k_{2}:=1 \TO  \ell (q[k_{1}]) \STEP  1 \DO 
       \IF  q[k_{1}][k_{2}].|wrtype| \neq ``R''
                \THEN  |wnet| := |wnet| - q[k_{1}][k_{2}].|wrqty|
                \ELSE  |wnet| := |wnet| + q[k_{1}][k_{2}].|wrqty| \FI ;
    k_{1} := k_{1}+ 1;
    |endgroup|() \OD ;
\end{program}
We can express the change to |wnet| as a function of the structure:
\begin{program}
\FUNCT  |change|(s) \EQ 
  \IF  s.|wrtype| \neq  ``R'' \THEN  -s.|wrqty| \ELSE  s.|wrqty| \FI \fullstop 
\end{program}
It is clear that the inner loop is computing the sum of the |change| outputs for all the structures in the sub
list $q[k_{1}]$, so we can collapse the inner loop to a reduce of a map operation:
\begin{program}
q := |split|(|input|, |same_item|);
k_{1}:= 1;
  \WHILE  k_{1} < \ell (q) \DO 
    |wnet| := {+} \bdiv  |change|\bstar q[k_{1}];
    k_{1} := k_{1}+ 1;
    |endgroup|() \OD ;
\end{program}
The |endgroup| procedure simply appends an element to the |output| list:
\begin{program}
q := |split|(|input|, |same_item|);
k_{1}:= 1;
  \WHILE  k_{1} < \ell (q) \DO 
    |wnet| := {+} \bdiv  |change|\bstar q[k_{1}];
    |output| := |output| \concat  \langle \langle q[k_{1}][1], |wnet|\rangle \rangle ;
    k_{1} := k_{1}+ 1;
    |endgroup|() \OD ;
\end{program}
so we can collapse the outer loop to a map operation. See Section~\ref{sec-spec} for the final
specification.

This extracted specification looks very different to the original assembler (see
Section~\ref{sec-assembler}) but both programs are semantically equivalent and generate identical
output files (when the output from the specification is formatted to match the assembler).

\section{Conclusion}

This paper describes a particularly challenging reverse engineering task: using formal program
transformations to extract a high-level abstract specification from an IBM 370 assembler program. The
original assembler program contains several ``layers'' of complexity including self-modifying code, a
flag used to direct control flow, a convoluted control flow structure and so on. Fortunately the powerful
automatic transformations implemented in \FermaT{} allow us to remove the first few layers of complexity
before we even have to look at the program. Moving to higher levels of abstraction requires a certain
amount of human intervention: particularly to select appropriate abstract data structures. However, this
intervention requires only localised analysis of the program. The higher-level control flow
transformations such as loop unrolling, loop rolling, taking code out of loops etc., are all implemented in
the \FermaT{} system and any global analysis required by these transformations is handled automatically.

\section{References}
\noindent\raggedright\small

\message{REFERENCE LIST}

\bgroup\Resetstrings%
\def\Loccittest{}\def\Abbtest{}\def\Capssmallcapstest{}\def\Edabbtest{}\def\Edcapsmallcapstest{}\def\Underlinetest{}%
\def\NoArev{0}\def\NoErev{0}\def\Acnt{1}\def\Ecnt{0}\def\acnt{0}\def\ecnt{0}%
\def\Ftest{ }\def\Fstr{1}%
\def\Atest{ }\def\Astr{R. J. R. Back}%
\def\Ttest{ }\def\Tstr{Correctness Preserving Program Refinements}%
\def\Stest{ }\def\Sstr{Mathematical Centre Tracts}%
\def\Vtest{ }\def\Vstr{131}%
\def\Itest{ }\def\Istr{Mathematisch Centrum}%
\def\Ctest{ }\def\Cstr{Amsterdam}%
\def\Dtest{ }\def\Dstr{1980}%
\Refformat\egroup%

\bgroup\Resetstrings%
\def\Loccittest{}\def\Abbtest{}\def\Capssmallcapstest{}\def\Edabbtest{}\def\Edcapsmallcapstest{}\def\Underlinetest{}%
\def\NoArev{0}\def\NoErev{0}\def\Acnt{4}\def\Ecnt{0}\def\acnt{0}\def\ecnt{0}%
\def\Ftest{ }\def\Fstr{2}%
\def\Atest{ }\def\Astr{F. L. Bauer%
  \Acomma B. Moller%
  \Acomma H. Partsch%
  \Aandd P. Pepper}%
\def\Ttest{ }\def\Tstr{Formal Construction by Transformation---Computer Aided Intuition Guided Programming}%
\def\Jtest{ }\def\Jstr{IEEE Trans. Software Eng.}%
\def\Vtest{ }\def\Vstr{15}%
\def\Ntest{ }\def\Nstr{2}%
\def\Dtest{ }\def\Dstr{Feb., 1989}%
\Refformat\egroup%

\bgroup\Resetstrings%
\def\Loccittest{}\def\Abbtest{}\def\Capssmallcapstest{}\def\Edabbtest{}\def\Edcapsmallcapstest{}\def\Underlinetest{}%
\def\NoArev{0}\def\NoErev{0}\def\Acnt{1}\def\Ecnt{0}\def\acnt{0}\def\ecnt{0}%
\def\Ftest{ }\def\Fstr{3}%
\def\Atest{ }\def\Astr{G. D. Bergland}%
\def\Ttest{ }\def\Tstr{A Guided Tour of Program Design Methodologies}%
\def\Jtest{ }\def\Jstr{Computer}%
\def\Itest{ }\def\Istr{IEEE}%
\def\Vtest{ }\def\Vstr{14}%
\def\Ntest{ }\def\Nstr{10}%
\def\Ptest{ }\def\Pcnt{ }\def\Pstr{18--37}%
\def\Dtest{ }\def\Dstr{Oct., 1981}%
\Refformat\egroup%

\bgroup\Resetstrings%
\def\Loccittest{}\def\Abbtest{}\def\Capssmallcapstest{}\def\Edabbtest{}\def\Edcapsmallcapstest{}\def\Underlinetest{}%
\def\NoArev{0}\def\NoErev{0}\def\Acnt{1}\def\Ecnt{0}\def\acnt{0}\def\ecnt{0}%
\def\Ftest{ }\def\Fstr{4}%
\def\Atest{ }\def\Astr{C. R. Karp}%
\def\Ttest{ }\def\Tstr{Languages with Expressions of Infinite Length}%
\def\Itest{ }\def\Istr{North-Holland}%
\def\Ctest{ }\def\Cstr{Amsterdam}%
\def\Dtest{ }\def\Dstr{1964}%
\Refformat\egroup%

\bgroup\Resetstrings%
\def\Loccittest{}\def\Abbtest{}\def\Capssmallcapstest{}\def\Edabbtest{}\def\Edcapsmallcapstest{}\def\Underlinetest{}%
\def\NoArev{0}\def\NoErev{0}\def\Acnt{1}\def\Ecnt{0}\def\acnt{0}\def\ecnt{0}%
\def\Ftest{ }\def\Fstr{5}%
\def\Atest{ }\def\Astr{C. C. Morgan}%
\def\Ttest{ }\def\Tstr{Programming from Specifications}%
\def\Itest{ }\def\Istr{Pren\-tice-Hall}%
\def\Ctest{ }\def\Cstr{Englewood Cliffs, NJ}%
\def\Otest{ }\def\Ostr{Second Edition}%
\def\Dtest{ }\def\Dstr{1994}%
\Refformat\egroup%

\bgroup\Resetstrings%
\def\Loccittest{}\def\Abbtest{}\def\Capssmallcapstest{}\def\Edabbtest{}\def\Edcapsmallcapstest{}\def\Underlinetest{}%
\def\NoArev{0}\def\NoErev{0}\def\Acnt{3}\def\Ecnt{0}\def\acnt{0}\def\ecnt{0}%
\def\Ftest{ }\def\Fstr{6}%
\def\Atest{ }\def\Astr{C. C. Morgan%
  \Acomma K. Robinson%
  \Aandd Paul Gardiner}%
\def\Ttest{ }\def\Tstr{On the Refinement Calculus}%
\def\Rtest{ }\def\Rstr{Technical Monograph PRG-70}%
\def\Itest{ }\def\Istr{Oxford University}%
\def\Dtest{ }\def\Dstr{Oct., 1988}%
\Refformat\egroup%

\bgroup\Resetstrings%
\def\Loccittest{}\def\Abbtest{}\def\Capssmallcapstest{}\def\Edabbtest{}\def\Edcapsmallcapstest{}\def\Underlinetest{}%
\def\NoArev{0}\def\NoErev{0}\def\Acnt{1}\def\Ecnt{0}\def\acnt{0}\def\ecnt{0}%
\def\Ftest{ }\def\Fstr{7}%
\def\Atest{ }\def\Astr{H. A. Partsch}%
\def\Ttest{ }\def\Tstr{Specification and Transformation of Programs}%
\def\Itest{ }\def\Istr{Springer-Verlag}%
\def\Ctest{ }\def\Cstr{New York--Heidelberg--Berlin}%
\def\Dtest{ }\def\Dstr{1990}%
\Refformat\egroup%

\bgroup\Resetstrings%
\def\Loccittest{}\def\Abbtest{}\def\Capssmallcapstest{}\def\Edabbtest{}\def\Edcapsmallcapstest{}\def\Underlinetest{}%
\def\NoArev{0}\def\NoErev{0}\def\Acnt{2}\def\Ecnt{0}\def\acnt{0}\def\ecnt{0}%
\def\Ftest{ }\def\Fstr{8}%
\def\Atest{ }\def\Astr{H. A. Priestley%
  \Aand M. Ward}%
\def\Ttest{ }\def\Tstr{A Multipurpose Backtracking Algorithm}%
\def\Jtest{ }\def\Jstr{J. Symb. Comput.}%
\def\Vtest{ }\def\Vstr{18}%
\def\Ptest{ }\def\Pcnt{ }\def\Pstr{1--40}%
\def\Dtest{ }\def\Dstr{1994}%
\def\Otest{ }\def\Ostr{\www{http:~//~www.~dur.~ac.~uk/~$\sim$dcs0mpw/~martin/~papers/~backtr-t.ps.gz}}%
\Refformat\egroup%

\bgroup\Resetstrings%
\def\Loccittest{}\def\Abbtest{}\def\Capssmallcapstest{}\def\Edabbtest{}\def\Edcapsmallcapstest{}\def\Underlinetest{}%
\def\NoArev{0}\def\NoErev{0}\def\Acnt{1}\def\Ecnt{0}\def\acnt{0}\def\ecnt{0}%
\def\Ftest{ }\def\Fstr{9}%
\def\Atest{ }\def\Astr{M. Ward}%
\def\Ttest{ }\def\Tstr{Proving Program Refinements and Transformations}%
\def\Rtest{ }\def\Rstr{DPhil Thesis}%
\def\Itest{ }\def\Istr{Oxford University}%
\def\Dtest{ }\def\Dstr{1989}%
\Refformat\egroup%

\bgroup\Resetstrings%
\def\Loccittest{}\def\Abbtest{}\def\Capssmallcapstest{}\def\Edabbtest{}\def\Edcapsmallcapstest{}\def\Underlinetest{}%
\def\NoArev{0}\def\NoErev{0}\def\Acnt{1}\def\Ecnt{0}\def\acnt{0}\def\ecnt{0}%
\def\Ftest{ }\def\Fstr{10}%
\def\Atest{ }\def\Astr{M. Ward}%
\def\Ttest{ }\def\Tstr{Program Analysis by Formal Transformation}%
\def\Jtest{ }\def\Jstr{Comput. J.}%
\def\Dtest{ }\def\Dstr{1996}%
\def\Vtest{ }\def\Vstr{39}%
\def\Ntest{ }\def\Nstr{7}%
\def\Dtest{ }\def\Dstr{1996}%
\def\Otest{ }\def\Ostr{\www{http:~//~www.~dur.~ac.~uk/~$\sim$dcs0mpw/~martin/~papers/~topsort-t.ps.gz}}%
\Refformat\egroup%

\bgroup\Resetstrings%
\def\Loccittest{}\def\Abbtest{}\def\Capssmallcapstest{}\def\Edabbtest{}\def\Edcapsmallcapstest{}\def\Underlinetest{}%
\def\NoArev{0}\def\NoErev{0}\def\Acnt{1}\def\Ecnt{0}\def\acnt{0}\def\ecnt{0}%
\def\Ftest{ }\def\Fstr{11}%
\def\Atest{ }\def\Astr{M. Ward}%
\def\Ttest{ }\def\Tstr{Assembler to C Migration using the FermaT Transformation System}%
\def\Jtest{ }\def\Jstr{International Conference on Software Maintenance, 30th Aug--3rd Sept 1999, Oxford, England}%
\def\Dtest{ }\def\Dstr{1999}%
\Refformat\egroup%

\bgroup\Resetstrings%
\def\Loccittest{}\def\Abbtest{}\def\Capssmallcapstest{}\def\Edabbtest{}\def\Edcapsmallcapstest{}\def\Underlinetest{}%
\def\NoArev{0}\def\NoErev{0}\def\Acnt{1}\def\Ecnt{0}\def\acnt{0}\def\ecnt{0}%
\def\Ftest{ }\def\Fstr{12}%
\def\Atest{ }\def\Astr{M. Ward}%
\def\Ttest{ }\def\Tstr{Abstracting a Specification from Code}%
\def\Jtest{ }\def\Jstr{J. Software Maintenance: Research and Practice}%
\def\Itest{ }\def\Istr{John Wiley \& Sons}%
\def\Ctest{ }\def\Cstr{New York, NY}%
\def\Vtest{ }\def\Vstr{5}%
\def\Ntest{ }\def\Nstr{2}%
\def\Ptest{ }\def\Pcnt{ }\def\Pstr{101--122}%
\def\Dtest{ }\def\Dstr{June, 1993}%
\def\Otest{ }\def\Ostr{\www{http:~//~www.~dur.~ac.~uk/~$\sim$dcs0mpw/~martin/~papers/~prog-spec.ps.gz}}%
\Refformat\egroup%

\bgroup\Resetstrings%
\def\Loccittest{}\def\Abbtest{}\def\Capssmallcapstest{}\def\Edabbtest{}\def\Edcapsmallcapstest{}\def\Underlinetest{}%
\def\NoArev{0}\def\NoErev{0}\def\Acnt{1}\def\Ecnt{0}\def\acnt{0}\def\ecnt{0}%
\def\Ftest{ }\def\Fstr{13}%
\def\Atest{ }\def\Astr{M. Ward }%
\def\Ttest{ }\def\Tstr{Derivation of Data Intensive Algorithms by Formal Transformation}%
\def\Jtest{ }\def\Jstr{IEEE Trans. Software Eng.}%
\def\Vtest{ }\def\Vstr{22}%
\def\Ntest{ }\def\Nstr{9}%
\def\Dtest{ }\def\Dstr{Sept., 1996}%
\def\Otest{ }\def\Ostr{\www{http:~//~www.~dur.~ac.~uk/~$\sim$dcs0mpw/~martin/~papers/~sw-alg.ps.gz}}%
\def\Ptest{ }\def\Pcnt{ }\def\Pstr{665--686}%
\Refformat\egroup%

\newpage
\section{The Assembler Source} \label{sec-assembler}

\label{sec-src}
\scriptsize
\begin{verbatim}
*****************************************
*  TST004A0  SAMPLE PROGRAM (MCDONALDS) *
*****************************************
*
         REGEQU
*
*        PRINT NOGEN
TST004A0 CSECT
         STM   R14,R12,12(R13)
         LR    R3,R15
         USING TST004A0,R3
         ST    R13,WSAVE+4
         LA    R14,WSAVE
         ST    R14,8(R13)
         LA    R13,WSAVE
*
         OPEN  (DDIN,(INPUT))
         OPEN  (RDSOUT,(OUTPUT))
*
         MVC   WPRT(17),=CL17'MANAGEMENT REPORT'
         BAL   R10,WRITE1
         BAL   R10,WRITE1
         MVC   WPRT(20),=CL20'ITEM      NET CHANGE'
         BAL   R10,WRITE1
         BAL   R10,WRITE1
*
         MVI   XSW1,0
LAA      EQU   *
         GET   DDIN,WREC
         CLC   WRITEM,WLAST
         BE    LAC
LAAA     B     LAB
         BAL   R10,ENDGROUP
LAB      MVI   LAAA+1,0
         MVC   WLAST,WRITEM
         ZAP   WNET,=P'0'
         BAL   R10,PROCGRP
         MVI   XSW1,X'FF'
         B     LAA
LAC      BAL   R10,PROCGRP
         MVI   XSW1,X'FF'
         B     LAA
*
LAD      CLI   XSW1,X'FF'
         BNE   LADA
         BAL   R10,ENDGROUP
LADA     EQU   *
         MVC   WPRT(17),=CL17'NUMBER CHANGED = '
         ED    WORKB,WCHANGE
         LA    R4,WORKB
         LA    R1,9
LADB     CLI   0(R4),C' '
         BNE   LADC
         LA    R4,1(R4)
         BCT   R1,LADB
LADC     EX    R1,WMVC1
*WMVC1   MVC   WPRT+17(1),0(R4)
         BAL   R10,WRITE1
*
         CLOSE DDIN
         CLOSE RDSOUT
*
         L     R13,WSAVE+4
         LM    R14,R12,12(R13)
         SLR   R15,R15
         BR    R14
*
PROCGRP  EQU   *
         ST    R10,WST10A
         PACK  WORKA,WRQTY
         CLI   WRTYPE,C'R'
         BNE   LBA
         AP    WNET,WORKA
         B     LBB
LBA      SP    WNET,WORKA
LBB      L     R10,WST10A
         BR    R10
*
ENDGROUP EQU   *
         ST    R10,WST10A
         MVC   WPRT(4),WLAST
         MVI   WSIGN,C'+'
         CP    WNET,=P'0'
         BNL   LCA
         MVI   WSIGN,C'-'
LCA      EQU   *
         MVC   WPRT+7(10),=X'40206B2020206B202120'
         EDMK  WPRT+7(10),WNET
         BCTR  R1,0
         MVC   0(1,R1),WSIGN
         BAL   R10,WRITE1
         BAL   R10,WRITE1
         AP    WCHANGE,=P'1'
         L     R10,WST10A
         BR    R10
*
WRITE1   EQU   *
         PUT   RDSOUT,WPRT
         MVC   WPRT,WSPACES
         BR    R10
*
WMVC1    MVC   WPRT+17(1),0(R4)
*
WSAVE    DC    18F'0'
WST10A   DS    F
WREC     DS    0CL80
WRITEM   DS    CL4
         DS    CL1
WRTYPE   DS    CL1
         DS    CL1
WRQTY    DS    CL3
         DS    CL70
WPRT     DC    CL80' '
WSPACES  DC    CL80' '
WLAST    DC    CL4'****'
WCHANGE  DC    PL4'0'
WNET     DC    PL4'0'
WORKA    DC    PL2'0'
WORKB    DC    XL10'40206B2020206B202120'
WSIGN    DC    CL1' '
XSW1     DC    X'00'
*
         LTORG
*
DDIN     DCB   DDNAME=DDIN,
               DSORG=PS,
               EODAD=LAD,
               MACRF=GM
RDSOUT   DCB   DDNAME=RDSOUT,
               DSORG=PS,
               MACRF=PM
*
         END
\end{verbatim}
\normalsize

\section{The WSL Specification}\label{sec-spec}

\vspace*{-\baselineskip}
\begin{program}
\BEGIN 
        q := |split|(|input|, |same_item|);
        |output| := \tab|header| \concat  |process|\bstar q
                                \quad \concat  \langle \langle ``NUMBER CHANGED = '', \ell (q)\rangle \rangle \untab
\WHERE 
        \FUNCT  |same_item|(x, y)\EQ  x.|writem| = y.|writem|\fullstop 
        \FUNCT  |process|(L)\EQ  \langle L[1], {+} \bdiv  |change|\bstar L\rangle \fullstop 
        \FUNCT  |change|(s)\EQ 
          \IF  s.|wrtype| \neq  ``R'' \THEN  -s.|wrqty| \ELSE  s.|wrqty| \FI \fullstop 
\END 
\end{program}

\end{document}